\documentclass[
 reprint,
 amssymb, amsmath,%
 groupedaddress
]{revtex4-1}

\usepackage{siunitx}
\usepackage{bm}%
\usepackage[colorlinks=true,linkcolor=blue]{hyperref}%
\usepackage{graphicx}%
\expandafter\ifx\csname package@font\endcsname\relax\else
 \expandafter\expandafter
 \expandafter\usepackage
 \expandafter\expandafter
 \expandafter{\csname package@font\endcsname}%
\fi
\begin{document}

\title{Measurement of the acoustic radiation force on a sphere embedded in a soft solid}%

\author{Pierre Lidon}%
\email{pierre.lidon@ens-lyon.org}
\author{Louis Villa}%
\author{Nicolas Taberlet}%
\author{S{\'e}bastien Manneville}%
\affiliation{Univ Lyon, Ens de Lyon, Univ Claude Bernard, CNRS, Laboratoire de physique, F-69342 Lyon, France}%

\date{November 2016}%
\revised{January 2016}%

\begin{abstract}
The acoustic radiation force exerted on a small sphere located at the focus of an ultrasonic beam is measured in a soft gel. It is proved to evolve quadratically with the local amplitude of the acoustic field. Strong oscillations of the local pressure are observed and attributed to an acoustic Fabry-P{\'e}rot effect between the ultrasonic emitter and the sphere. Taking this effect into account with a simple model, a quantitative link between the radiation force and the acoustic pressure is proposed and compared to theoretical predictions in the absence of dissipation. The discrepancy between experiment and theory suggests that dissipative effects should be taken into account for fully modeling the observations. 
\end{abstract}

\maketitle



\textit{Introduction -} Acoustic manipulation has recently emerged as a versatile tool to displace, sort and organize small objects such as droplets, bubbles or cells in microfluidic devices \cite{petersson_2004,wiklund_2006,evander_2007,rabaud_2011}. Indeed, when submitted to an acoustic wave, the interface between two materials displaying an acoustic contrast is exposed to a force $F_\text{rad}$, known as the acoustic radiation force, that arises from non-linear acoustic effects\cite{chen_1996}. According to the sign of the acoustic contrast, it is possible to push or pull objects in a fluid with a traveling wave\cite{hertz_1939,schmid_2014,li_2015a} or to induce the migration of particles towards the nodes or the antinodes of a standing wave \cite{goldman_1949,tran_2012,guo_2016}. Thus $F_\text{rad}$ allows one to move precisely small objects\cite{laurell}, which has opened a path for microfluidic design of structured materials via bottom-up approaches\cite{chen_2013,gesellchen_2014,ma_2015,caleap_2014,caleap_2016}. In order to preserve the obtained mesoscopic structure after insonification, it has been proposed to work in photocurable resins\cite{raeymaekers_2011} or in yield-stress fluids\cite{xu_2011}. However, while the acoustic radiation force is well characterized in simple inviscid fluids, the effect of viscosity on $F_\text{rad}$ is still under debate\cite{doinikov_1994a,settnes_2012,karlsen_2015,sepehrirahnama_2015,karlsen_2016,sepehrirahnama_2016} and only a few studies have been devoted to the general case of viscoelastic media\cite{chen_2002,ilinskii_2005,aglyamov_2007,urban_2011,suomi_2015}. It is thus of growing interest to characterize the acoustic radiation force in complex, soft materials.

In this Letter, we show experimentally that the acoustic radiation force exerted on a rigid sphere embedded in a viscoelastic solid and submitted to a focused ultrasonic field is related to the local pressure $P_\text{loc}$ by $F_\text{rad} = \beta P_\text{loc}^2$ and we provide an estimate of the parameter $\beta$. We further show that there exists a partially standing wave between the sphere and the acoustic transducer, so that $F_\text{rad}$ is highly sensitive to the sphere position. This cavity effect is well captured by a simple model. Finally, the measured prefactor $\beta$ is found to be larger by about 40\% than the theoretical value in the absence of dissipation, which hints at the necessity to account for viscosity in order to correctly predict $F_\text{rad}$ in soft materials.



\begin{figure}[!htb]
\centering
\includegraphics[scale=1]{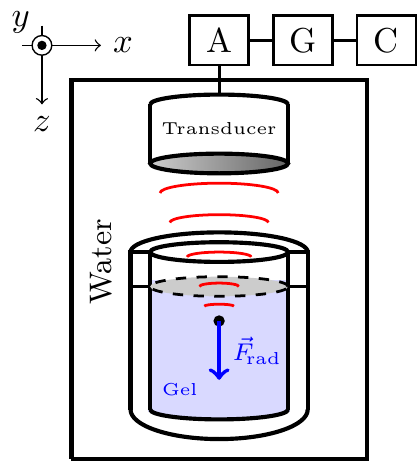}
\caption{Experimental setup. The cell and the transducer are immersed in a water bath. ``A'' stands for power amplifier, ``G'' for function generator and ``C'' for computer. \label{fig:transducteur_setup}}
\end{figure}

\textit{Principle of the experiment -} In order to quantify $F_\text{rad}$, the motion of an intruder within a soft gel under the effect of focused ultrasound is recorded, as sketched in Fig.~\ref{fig:transducteur_setup}. A similar experiment has been used to assess the mechanical properties of tissue-like gelatin gels by analyzing the transient displacement of the intruder induced by short pulses of duration up to a few milliseconds\cite{aglyamov_2007}. Here, we propose a different approach: assuming that the rheology of the gel is known, we measure the acoustic radiation force through the steady-state displacement $\delta r_\text{max}$ of a spherical intruder exposed to much longer insonification. Indeed, assuming a no-slip boundary condition at the sphere surface, $F_\text{rad}$ is related to $\delta r_\text{max}$ by\cite{urban_2011,ilinskii_2005}
\begin{equation}
F_\text{rad} = 6\pi G_0 a \delta r_\text{max}\, ,
\label{eq:link_displacement_force}
\end{equation}
\noindent where $G_0$ is the elastic modulus of the material and $a$ the radius of the spherical intruder.
 
\textit{Setup -} This study focuses on a gel made of carbopol ETD 2050 dispersed in water at a concentration of $1\%$~wt and prepared following a protocol already described elsewhere \cite{lidon_2016b}. Such a gel has a yield stress $\sigma_\text{y}=\SI{19.5}{\pascal}$ below which it behaves as a soft viscoelastic solid of elastic modulus $G_0 \simeq \SI{55}{\pascal}$ and notably smaller viscous modulus ($\simeq\SI{4}{\pascal}$) as measured by a stress-controlled rheometer (Anton Paar MCR-301). The gel is first centrifugated to remove air bubbles. It is then progressively poured into a cylindrical cell of inner diameter $\SI{2}{\centi\meter}$ and length $\SI{5}{\centi\meter}$. A single polystyrene sphere of radius $a=\SI{163 \pm 3}{\micro\meter}$ is introduced into the gel. As the buoyancy stress due to the density mismatch between the sphere ($\rho_\text{p} \simeq \SI{1.05}{\kilo\gram\per\meter\cubed}$) and the gel ($\rho_\text{g} \simeq \SI{1.03}{\kilo\gram\per\meter\cubed}$) is smaller than $\sigma_\text{y}$, the sphere does not sediment. Finally, the cell is closed with a thin plastic wrap. During all these steps, great care is given not to trap any air bubbles within the gel, which would result in important distortion of the acoustic waves.

The cell is moved using a three-axis translation stage controlled by a computer. The setup is illuminated by a LED panel providing a uniform luminous background. Images of the sphere are recorded at a rate of $300 \, \text{fps}$ with a CCD camera (Baumer HXC20) mounted on a macroscope (Nikon SMZ745T). Images are analyzed with a standard particle detection algorithm allowing us to measure the sphere trajectory with a resolution of about $\SI{1}{\micro\meter}$.

Finally, ultrasonic waves are produced by a focused piezoelectric transducer (Imasonic, diameter $\SI{38}{\milli\meter}$, center frequency $f=\SI{2.25}{\mega\hertz}$). The cell and the transducer are placed within a water bath at temperature $T \simeq \SI{20}{\celsius}$ in order to ensure ultrasound propagation. We checked that the gel, the plastic membrane and the cell do not distort nor attenuate the ultrasonic field. The transducer is driven by a power amplifier (Kalmus 150C) controlled by a function generator (Agilent 33522A). It emits bursts of duration $\SI{0.5}{\second}$, hereafter referred to as ``pulses,'' and it is left to rest for about one minute to allow for full relaxation of the gel. The sphere can be located precisely at the focus of the transducer by maximizing its displacement under a fixed acoustic intensity.

\begin{figure}[!htb]
\centering
\includegraphics[scale=1]{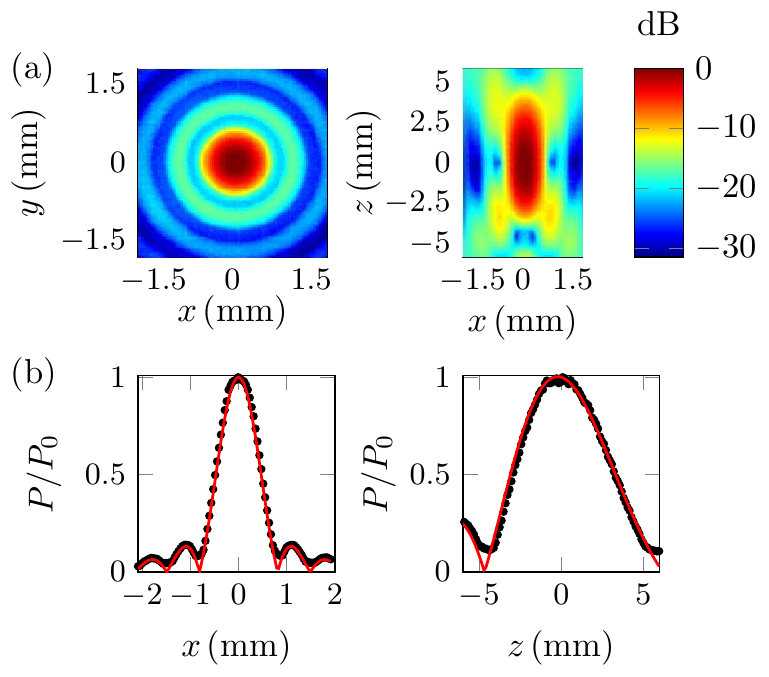}
\caption{(a) Pressure field in water and in the absence of a sphere, around the focal spot of the piezoelectric transducer in the transverse plane (left) and in the propagation plane (right). The color scale corresponds to relative acoustic intensity in decibels, i.e. to $20\log{(P/P_0)}$. (b) Pressure profiles $P$ at focus along a perpendicular axis $x$ (left) and along the direction of propagation $z$ (right), normalized by the maximum pressure at focus $P_0$. Red curves are theoretical predictions by Eqs.~(\ref{eq:pressure_focused_transducer_x}) and~(\ref{eq:pressure_focused_transducer_z}) with no adjustable parameter. \label{fig:transducteur_champ}}
\end{figure}

Figure~\ref{fig:transducteur_champ}(a) shows the pressure field characterization $P(x,y,z)$ obtained by scanning the vicinity of the focal spot with a needle hydrophone (Precision Acoustics 1717SN with active element diameter $\SI{0.2}{\milli\meter}$) in water and in the absence of the experimental cell. The acoustic field is axisymmetric about the direction of propagation and pressure profiles are displayed in Fig.~\ref{fig:transducteur_champ}(b). They are well fitted by the following theoretical expressions without any adjustable parameter\cite{kino}:
\begin{align}
P(x,0,0) = &P_0 \frac{\lambda \ell}{\pi a x} \mathrm{J}_1 \left(\frac{2\pi a x}{\lambda \ell} \right) \label{eq:pressure_focused_transducer_x} \\
P(0,0,z) =  &P_0 \frac{\ell}{z+\ell} \mathrm{sinc} \left(\frac{a^2}{2\lambda \ell} \frac{z}{z+\ell} \right)
\label{eq:pressure_focused_transducer_z}
\end{align}
\noindent where $J_1$ is the Bessel function of the first kind, $a=\SI{19}{\milli\meter}$ is the radius of the transducer, $\ell=\SI{38}{\milli\meter}$ its focal length and $\lambda = c_\text{w}/f=\SI{670}{\micro\meter}$ is the acoustic wavelength in water for the considered frequency and sound speed $c_\text{w}=\SI{1480}{\meter\per\second}$ at room temperature. The acoustic field is focused in a small spot of diameter $\SI{1.0}{\milli\meter}$ and focal depth at $\SI{-3}{\deci\bel}$ $\ell_\text{z}\simeq\SI{5}{\milli\meter}$. Consequently, it can be assumed to be homogeneous at the scale of the sphere so that $F_\text{rad}$ will be estimated locally at the sphere position.



\begin{figure}[!htb]
\centering
\includegraphics[scale=1]{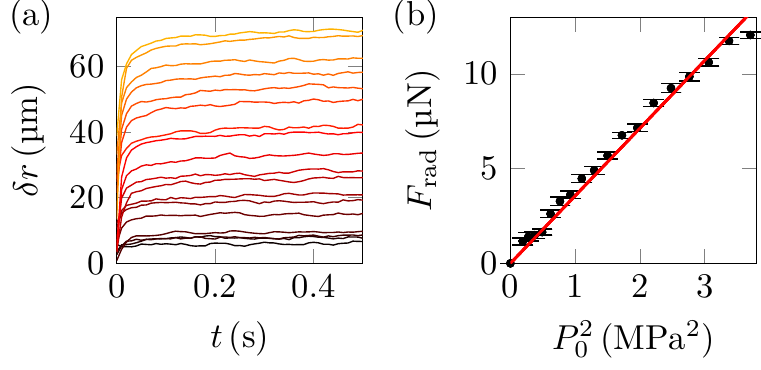}
\caption{(a) Displacement $\delta r$ of the sphere as a function of time $t$ during pulses of duration $\SI{0.5}{\second}$ and amplitude ranging from $P_0=\SI{0.4}{\mega\pascal}$ to $P_0=\SI{2.0}{\mega\pascal}$. (b) Radiation force $F_\text{rad}$ obtained from Eq.~\eqref{eq:link_displacement_force} as a function of $P_0^2$. The red line is a linear fit, $F_\text{rad} = \alpha P_0^2$, with $\alpha=\SI{3.7(3)e-18}{\newton\per\pascal\squared}$.  \label{fig:creep}}
\end{figure}

\textit{Scaling of $F_\text{rad}$ with acoustic intensity -} Figure~\ref{fig:creep}(a) displays the sphere displacement $\delta r$ during successive pulses of increasing amplitude $P_0=\SI{0.4}{\mega\pascal}$ to $\SI{2.0}{\mega\pascal}$. The initial position of the sphere corresponds to the focus of the transducer and the maximum value of $P_0$ is chosen such that the acoustic radiation pressure remains below the yield stress in order to avoid any irrecoverable displacement of the sphere.

For the range of acoustic intensities considered here, the sphere reaches an almost stationary displacement $\delta r_\text{max}$ at the end of the pulse and $\delta r_\text{max}$ is converted into an estimate of $F_\text{rad}$ by using Eq.~\eqref{eq:link_displacement_force}. As shown in Fig.~\ref{fig:creep}(b), the acoustic radiation force scales as $P_0^2$, with a slope $\alpha=\SI{3.7(3)e-18}{\newton\per\pascal\squared}$. Although such a quadratic scaling is not surprising --since $F_\text{rad}$ results from a non-linear effect\cite{chen_1996}--, it is nicely confirmed here for a viscoelastic propagation medium.



\begin{figure}[!htb]
\centering
\includegraphics[scale=1]{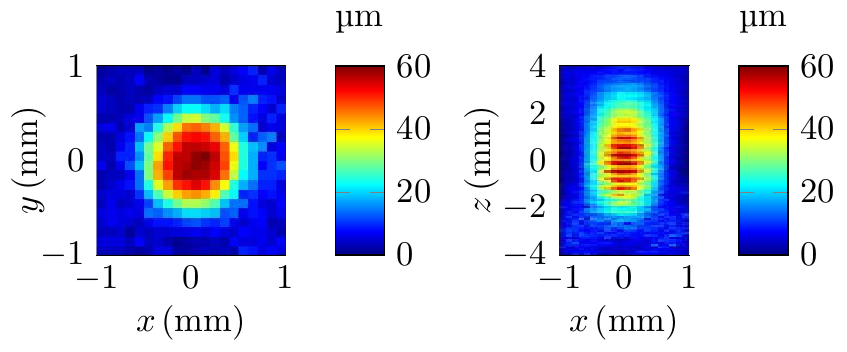}
\caption{Maximum displacement $\delta r_\text{max}$ of the sphere during pulses of duration $\SI{0.5}{\second}$ and amplitude $P_0=\SI{1.8}{\mega\pascal}$ for different initial positions across the pressure field, in the transverse plane (left) and in the propagation plane (right). \label{fig:champ_force}}
\end{figure}

\textit{Oscillations of $F_\text{rad}$ with sphere position -} In order to explore the effect of the sphere position with respect to the transducer on $F_\text{rad}$, the focal area is scanned by moving the cell in horizontal $(xOy)$ and vertical $(xOz)$ planes centered around the focus by steps of size $\SI{100}{\micro\meter}$. At each position, $\delta r_\text{max}$ is recorded for a fixed pressure at focus $P_0=\SI{1.8}{\mega\pascal}$ as displayed in Fig.~\ref{fig:champ_force}.
It should be noted that the displacement map has the same overall shape and size as $P^2(x,y,z)$. However, striking oscillations are observed in the direction of ultrasound propagation $(Oz)$.

\begin{figure}[!htb]
\centering
\includegraphics[scale=1]{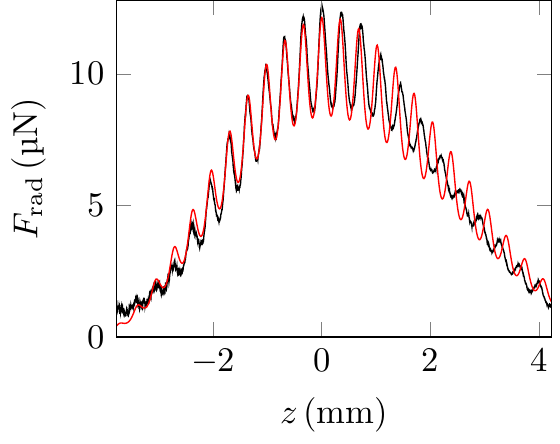}
\caption{Acoustic radiation force $F_\text{rad}$ exerted on the sphere as a function of the distance $z$ from focus along the direction of propagation for $P_0=\SI{1.8}{\mega\pascal}$. The red curve corresponds to Eqs.~\eqref{eq:model_cavity1} and \eqref{eq:model_cavity2} with $r=-0.115$, $\rho=-0.8$ and $\lambda=\SI{680}{\micro\meter}$. \label{fig:cavite_exp}}
\end{figure}

\begin{figure}[!htb]
\centering
\includegraphics[scale=1]{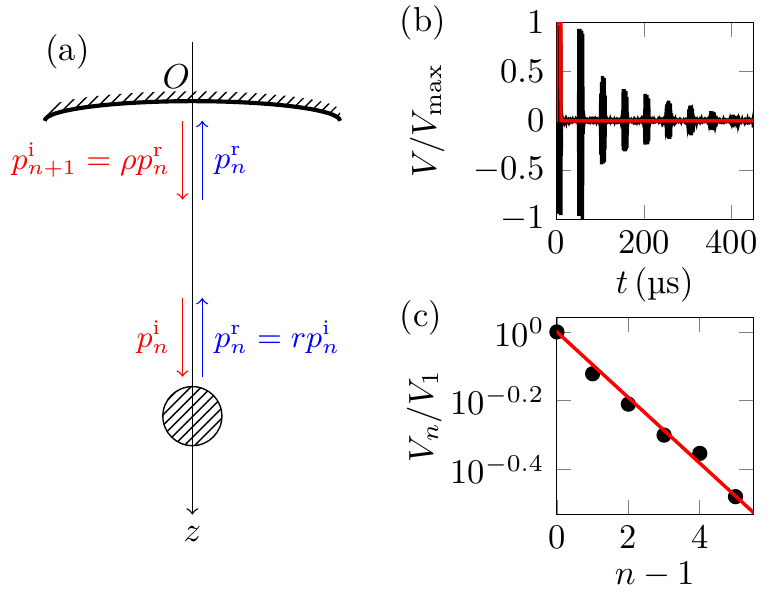}
\caption{(a) Schematics of the cavity effect: the acoustic wave is successively reflected by the sphere and by the transducer, leading to various pressure contributions $p_n^{\rm i}$ and $p_n^{\rm r}$ that add up and form a partially standing wave in the cavity. (b) Voltage output of the transducer after successive reflections between the transducer and a water--air interface. The red curve corresponds to the electrical input of duration $\SI{10}{\micro\second}$. (c) Amplitude of the successive echoes as a function of the number of reflections $n$. The red line shows the linear fit $\log{(V_n/V_1)}= (n-1)\log{|\rho|}$ with $\rho=-0.80$. \label{fig:cavite_modele}}
\end{figure}

In order to better characterize these oscillations, Fig.~\ref{fig:cavite_exp} shows a scan performed along the line $(Oz)$ with smaller steps of $\SI{2}{\micro\meter}$. The oscillations display a period close to $\lambda / 2$ and are sharper on the upper part. These features are reminiscent of a Fabry-P{\'e}rot or cavity effect. During the pulse, acoustic waves indeed are reflected back and forth on the sphere and on the transducer a large number of times. This results in the formation of an acoustic cavity: the acoustic wave between the sphere and the transducer is the superposition of the incident traveling wave generated by the transducer and of the standing wave due to the Fabry-P{\'e}rot effect. This partially standing wave structure leads to oscillations of the acoustic force when the sphere is displaced with respect to the transducer\cite{issenmann_2006}.
This effect can be accounted for by a simple model sketched in Fig.~\ref{fig:cavite_modele}(a). We consider the sphere and the transducer as plane reflectors of reflection coefficients in amplitude $r$ and $\rho$ respectively. For a plane wave, computing the local pressure field $P_\text{loc}$ in the cavity from $r$ and $\rho$ is straightforward. Assuming that the case of a focused beam is simply obtained by modulating this field by the pressure profile $P(0,0,z)$ given by Eq.~\eqref{eq:pressure_focused_transducer_z}, we get 
\begin{eqnarray}
F_\text{rad}(z) &=& \alpha \Phi(z) P(0,0,z)^2\label{eq:model_cavity1}\\
\hbox{\rm{with~}}\Phi(z)&=&\frac{A+B \cos{[2k(z+\ell)]}}{\left(C + D \cos{[2k(z+\ell)]} \right)^2},
\label{eq:model_cavity2}
\end{eqnarray}
\noindent where $k=2\pi / \lambda$ and $A$, $B$, $C$ and $D$ depend only on $r$ and $\rho$\cite{expression_ABCD}.

The speed of sound in our carbopol gel was measured to be $c_\text{g}=\SI{1530}{\meter\per\second}$, corresponding to $\lambda=c_\text{g}/f=\SI{680}{\micro\meter}$. We also estimated the reflection coefficient $\rho\simeq -0.8$ by analyzing the decay of successive echoes of short bursts reflected on the transducer and on a flat, perfectly reflecting water--air interface, as shown in Fig.~\ref{fig:cavite_modele}(b,c). Finally, with the value of $\alpha$ inferred from the previous set of experiments, we obtain a good agreement between the model and experimental data by taking $r=-0.115$ as the sole free parameter (see red curve in Fig.~\ref{fig:cavite_exp}). This value of $r$ should only be interpreted as an effective reflection coefficient for the gel--sphere interface since the sphere diameter is comparable to the acoustic wavelength and a diffraction-based model would be required to fully grasp the details of our focused beam impinging on a sphere smaller than $\lambda$. Moreover, we observe that the oscillation period of the force increases past the focal spot, for $z>0$. This effect, which is not captured by the model, is too significant to be caused by a global temperature drift during the scan along $(Oz)$. It could result from a shift in the spectral content to lower frequencies due to nonlinear propagation and/or attenuation, although further work should confirm such an explanation.

\textit{Discussion -} 
Due to the aforementioned cavity effect, the local pressure field $P_\text{loc}$ varies significantly on a typical scale $\lambda/2 \simeq \SI{350}{\micro\meter}$ so that, for sphere displacements larger than about $\SI{100}{\micro\meter}$, $F_\text{rad}$ cannot be considered as a constant. Therefore, accurate measurements of the acoustic radiation force require both a good control of the sphere position and moderate displacements, as in Fig.~\ref{fig:creep}(a) for which the sphere was carefully located at focus and $\delta r$ did not exceed $\SI{70}{\micro\meter}$.

Moreover, $P_\text{loc}$ differs from the bare pressure field generated by the transducer in the absence of the sphere through a modulation by the oscillating function $\Phi$. Thus, in order to recover the ``true'' prefactor $\beta$ linking $F_\text{rad}$ and $P_\text{loc}^2$, the coefficient $\alpha$ measured in Fig.~\ref{fig:creep}(b) must be corrected for the effect of the cavity: at focus, one has $P_\text{loc}^2=\Phi(z=0) P_0^2$ so that $\beta = \alpha / \Phi(z=0)$, which yields $\beta =  \SI{7.5(7)e-18}{\meter\squared\per\pascal}$.

In the literature, the prefactor $\beta$ has been calculated analytically for an elastic sphere located at the center of a focused acoustic field and suspended in an inviscid fluid\cite{chen_1996}. These calculations should also apply to a suspending medium made of an isotropic elastic solid since the propagation equations remain unchanged as long as there is no dissipation.  They predict $\beta_\text{th} = \pi a^2 Y / 4\rho_\text{g} c_\text{g}^2$, where $Y$ involves the ratio $a/\lambda$, the densities ($\rho_\text{p}$, $\rho_\text{g}$) and sound speeds ($c_\text{p}$, $c_\text{g}$) of the particle and gel respectively [see Eq.~(63) in Ref.~\cite{chen_1996} for the full expression of $Y$]. Using the tabulated value for $c_\text{p}$ in polystyrene, we find $\beta_\text{th} = \SI{5.3e-18}{\meter\squared\per\pascal}$, which is of the same order of magnitude as our measurement.

The measured $\beta$ is, however, larger than the theoretical prediction by about 40\%, suggesting that the theory underestimates $F_\text{rad}$. This discrepancy could be attributed to dissipative effects in the carbopol gel which have been neglected here, both in the theory and in Eq.~\eqref{eq:link_displacement_force}. For instance, in Newtonian fluids, viscosity has been shown to increase the amplitude of the radiation force\cite{doinikov_1994a,settnes_2012,karlsen_2015,sepehrirahnama_2015} which could account for an underestimation of $\beta_\text{th}$. Thus, further theoretical work is needed to incorporate dissipative effects both in acoustical and rheological models.



\textit{Conclusion and perspectives -} To summarize, we have measured the acoustic radiation force exerted by a focused acoustic beam on a spherical obstacle embedded in a soft gel and we found that the acoustic force scales quadratically with the local pressure. Due to a cavity effect, the local pressure oscillates significantly when the distance between the sphere and the transducer is varied. This  implies that an accurate measurement of $F_\text{rad}$ requires a precise positioning of the sphere.

The setup described here is not restricted to the case of yield-stress fluids: provided that buoyant motion is slow enough compared to the experiment duration, this method can be employed for any viscoelastic material. Future work shall thus focus on measuring the acoustic radiation pressure for different intruders, surrounding fluids and pressure fields. To this aim, we will use a microscope to resolve smaller displacements and improve accuracy of the sphere positioning. We will moreover use higher acquisition rates to measure and analyze transient displacements with a higher resolution. This will provide data sets to characterize acoustic radiation pressure in complex materials in situations relevant for microfluidics but also for ultrasonic imaging and therapy. To rationalize the observations, this experimental effort will need to be accompanied by numerical and theoretical modeling of acoustic forces in complex media.



\textit{Ackowledgements -} This work was funded by the Institut Universitaire de France and by the European Research Council under the European Union's Seventh Framework Programme (FP7/2007-2013)/ERC grant agreement No. 258803.

The authors are grateful to Y. Forterre, G. Ovarlez and R. Wunenburger for fruitful discussions.

\newpage

\onecolumngrid

\begin{center}
\Large Erratum: ``Measurement of the acoustic radiation force on a sphere embedded in a soft solid'' [APL 110, 044103 (2017)]
\end{center}

In this erratum, we make corrections to our recent publication [APL 110, 044103 (2017)].

First, the densities of polystyrene and gel are in the wrong unit and should read $\rho_\text{p} \simeq \SI{1.05e3}{\kilo\gram\per\meter\cubed}$ and $\rho_\text{g} \simeq \SI{1.03e3}{\kilo\gram\per\meter\cubed}$.

Second, Eq.~(4) should be written as follows:
\begin{equation}
F_\text{rad}(z) = \alpha \Phi(z) P(0,0,z)^2 .
\end{equation}
\noindent This correction does not affect the rest of the article.

Third and most importantly, the theoretical value of $\beta$ is wrong by a factor $2$. The model of Ref.~5 indeed predicts $\beta_\text{th} = \pi a^2 Y / 2 \rho_\text{g} c_\text{g}^2$ and thus $\beta_\text{th} = \SI{10.6e-18}{\meter\squared\per\pascal}$, which is twice the value that is given in the article. This does not change our conclusion that the model fails at describing our measurements but it actually overestimates the experimental value of $\beta$ by about $\SI{40}{\percent}$. This invalidates our short subsequent discussion (top of page 4) that invoked viscous effects to justify that $\beta_\text{th}<\beta$.


\textit{Ackowledgements -} The authors are grateful to G.T.~Silva for bringing these mistakes to their attention.

\end{document}